\documentclass[a4paper,notitlepage,12pt]{article}

\begin{document}
\title{Terminal states of thermocapillary migration of a planar droplet
at moderate and large Marangoni numbers}
\author{ Zuo-Bing Wu$^{1,2}$ \footnotemark[1]\\
 $^1$State Key Laboratory of Nonlinear Mechanics,
 Institute of Mechanics,\\ Chinese Academy of Sciences,
 Beijing 100190, China\\
$^2$School of Engineering Science, \\University of Chinese Academy of Sciences,
  Beijing 100049, China}
 \maketitle

\footnotetext[1]{Correspondence to: wuzb@lnm.imech.ac.cn.}

\newpage
\begin{abstract}
In this paper, thermocapillary migration of a planar droplet at moderate and large Marangoni numbers is investigated analytically and numerically.
By using the dimension-analysis method, the thermal diffusion time scale is determined as the controlling one of
 the thermocapillary droplet migration system. During this time, the whole thermocapillary migration process is fully developed.
By using the front-tracking method,
 the steady/unsteady states as the terminal ones at moderate/large Marangoni numbers are captured
 in a longer time scale than the thermal diffusion time scale.
 In the terminal states, the instantaneous velocity fields in the unsteady migration process at large Marangoni numbers
 have the forms of the steady ones at moderate Marangoni numbers.
 However, in view of the former instantaneous temperature fields, the surface tension of the top surface of the droplet
 gradually becomes the main component of the driving force on the droplet after the inflection point appears.
 It is different from that the surface tension
 of the bottom surface of the droplet is the main component of the driving force on the droplet for the latter ones.
 The physical mechanism of thermocapillary droplet migration can be described as
 the significance of the thermal
 convection around the droplet is higher than/just as the thermal conduction across the droplet at large/moderate Marangoni numbers.


\end{abstract}

\newpage
\section{Introduction}
An immiscible droplet or bubble is placed in an ambient fluid with
temperature gradient in microgravity environment, it will move in
the direction of increasing temperature due to
the surface tension. This phenomenon is called as thermocapillary migration
in fundamental hydrodynamics and has many practical applications[1].
The pioneering work in [2] used a linear model to predict
 the migration speed of a droplet in the limits of zero Reynolds (Re) and
Marangoni (Ma) numbers. To include inertial
effects, the above analysis was extended to treat
the weak nonlinear model in the range of small Ma numbers[3].
Under the quasi-steady state assumption, the analytical results
were confirmed with the experimental results[4].
Since then, thermocapillary migration of a droplet in a large range
of Ma numbers has been studied
extensively by a series of the theoretical analyses[5],
numerical simulations[6] and experimental investigations[7].
In particular, for large Ma numbers, it was reported
that the migration speed of a droplet increases with increasing Ma
number[8]. The theoretical results are in a qualitative agreement
with the correspondent numerical simulations[9],
but are qualitatively inconsistent with the experimental investigations[10,11].
In the above works for large Ma numbers, both the theoretical analyses and
numerical simulations are based on the assumption of quasi-steady
state. However, in the experiment
investigations the droplet migrations are in an accelerating process
and do not reach steady states.
Therefore, although the thermocapillary droplet migration at small
Ma numbers is well understood, it is unclear what kinds of states might happen
in the thermocapillary droplet migration at large Ma numbers.
Extension of the quasi-steady state assumption proposed at small Ma numbers
to large Ma numbers is still a topic to be
studied with emphasis laid on the physical mechanism.
Moreover, in view of the mechanism of the varied surface tension with temperature,
the thermocapillary migration of a droplet was extended
to drive a film or a droplet on a substrate with a horizontal temperature
gradient, such as the examples
of films climbing planes by thermocapillary effect\cite{a1,a7,a2}
and moving droplets on horizontal substrates\cite{a3,a4,a5,a6}.
This subject called as thermocapillary actuation has
the similar physical mechanism as thermocapillary droplet migration and
potential applications on the chemical industry and biological engineering.

To investigate thermocapillary droplet migration
in a large range of Ma numbers, steady-unsteady thermocapillary
migrations of a planar droplet at moderate and large
Ma numbers were observed in both the numerical and analytical studies\cite{12},
where the numerical results are in qualitative agreement with the
experimental results.
However, the controlling time scale of
whole migration process is the thermal diffusion time scale and far greater
than the convective one.
It is expected that a steady state only appears in the relative time scale
of the order $O(Ma)$ when the convective time scale is adopted to non-dimensionalize time.
In this paper, using the dimension-analysis method, we first determine four different time scales
in the thermocapillary migration system and connect them with the non-dimensional
parameters of the system. Then,
using the front-tracking method, we numerically study
thermocapillary migration of a planar droplet at moderate and large Ma numbers and
capture their terminal states in a longer time scale,
which is beyond the relative thermal diffusion time scale $O(Ma)$.
The physical mechanism of the steady/unsteady migration at the moderate/large
Ma numbers is further analyzed by comparing the time variation of the
temperature fields at the front and the rear of the droplet and
determining the driving force on the droplet based on the terminal temperature
distributions at the interface of the droplet.

The paper is organized as follows. In section 2, we present
the governing equations of the problem of thermocapillary migration of a planar droplet
 and the numerical methods to solve them.
In section 3, we make the dimension analysis of the above problem to obtain
the controlling time scale of the whole thermocapillary migration process.
In section 4, numerical results for two cases of temperature gradients
$G=12$K/cm and 9 K/cm are analyzed.
 Finally, in section 5, some conclusions and discussions are given.

\section{Governing equations and numerical methods}

Consider the thermocapillary migration of a planar droplet
with the radius $R_0$ in a
continuous phase fluid of infinite extent (with the dynamical viscosity $\mu_1$)
 under a uniform
temperature gradient $G$.
The reference velocity is defined as
\begin{equation}
v_o=-\sigma_\Theta G R_0/\mu_1,
\end{equation}
where  $\sigma_\Theta(=d\sigma/d\Theta)$ is the rate of change of surface
tension with temperature.
 By taking $R_0$, $v_o$ and $GR_0$ as characteristic quantities to make coordinates,
velocity and temperature dimensionless,
the continuity, momentum
and energy equations for the continuous phase fluid and the
droplet in a laboratory coordinate system are written in
the following non-dimensional form
\begin{equation}
\begin{array}{l}
\frac{\partial{\rho_i}}{\partial t} + \nabla \cdot (\rho_i {\bf v}_i)=0,\\
\frac{\partial\rho_i {\bf v}_i}{\partial t} + \nabla \cdot (\rho_i
{\bf v}_i{\bf v}_i) = - \nabla p_i + \frac{1}{Re} \nabla \cdot
\mu_i(\nabla
{\bf v}_i +\nabla {\bf v}_i^T) + {\bf f}_{\sigma},\\
\frac{\partial \Theta_i}{\partial t} + \nabla \cdot ({\bf v}_i \Theta_i) =
\frac{1}{Ma} \frac{\kappa_i}{k_i} \nabla \cdot ({k_i \nabla \Theta_i}),
\end{array}
\end{equation}
where ${\bf v}_i=(u_i,v_i)$, $p_i$ and $\Theta_i$ are velocity, pressure and temperature,
respectively.
The physical coefficients (density $\rho_i$, dynamic viscosity $\mu_i$, thermal
conductivity $k_i$ and thermal diffusivity $\kappa_i$) are non-dimensionlized by
the quantities of continuous fluid.
Symbols with subscript 1
and 2 denote physical coefficients of the continuous fluid and the
droplet, respectively.
 ${\bf f}_{\sigma}$ is the scaled surface tension
acting on the interface.
 The Re and Ma
numbers are respectively defined as
\begin{equation}
 Re=\frac{\rho_1 v_0 R_0}{\mu_1},\ \ \  Ma=\frac{v_0R_0}{\kappa_1}.
\end{equation}
As shown schematically in Fig. 1, only half of the velocity/temperature field is determined due
to the mirror symmetry about $z$ axis involved in the system.
The solutions of Eqs. (2) satisfy the following initial
conditions of the half domain $x \in [0,x_1]$ and $z \in
[z_0,z_1]$
\begin{equation}
{\bf v}_i =0, \ \ \ \Theta_i=z
\end{equation}
and boundary conditions at the top and bottom
walls ($z=z_1$ and $z=z_0$), on the central symmetric axis ($x=0$) and at the right wall ($x=x_1$)
\begin{equation}
\begin{array}{l}
{\bf v}_1(x,z_0) ={\bf v}_1(x,z_1) =0, \Theta_1(x,z_0) =z_0, \Theta_1(x,z_1) =z_1,\\
u_i(0,z)=0, \frac{\partial v_i}{\partial x}(0,z)=0, \frac{\partial{\Theta_i}}{\partial x}(0,z)=0,\\
 {\bf v}_1(x_1,z)=0, \Theta_1(x_1,z)=z.
\end{array}
\end{equation}

To simulate the thermocapillary droplet migration, a fixed regular staggered MAC grid in
the computational domain is used.  For discretizing Eqs. (2),
 a second-order central difference scheme
for the spatial variables and
an explicit predictor-corrector second-order scheme for time
integration are adopted.
By using the front-tracking method\cite{13}, the immiscible interface is
considered to have a finite width, so that all physical
coefficients across the interface are continuous. Here, a weighting function\cite{14} is taken as
 \begin{equation}
 w_{ij}({\bf r}_p) =d(x_p-i\Delta x)d(z_p-j\Delta z),
 \end{equation}
where

\begin{equation}
 d(r) = \left \{ \begin{array}{ll}
 (1/4 \Delta r)[1 +\cos (\pi r/2 \Delta r)], &|r|<2 \Delta r,\\
  0,                                 &|r| \ge 2 \Delta r,
                 \end{array} \right.
\end{equation}
and $(x_p, z_p)$ is the interface node.
With the updated physical coefficients, the velocity, pressure and temperature fields
are computed by the Chorin's projection method.
Meanwhile, the non-dimensional surface tension\cite{13} is written in the form of body force as
\begin{equation}
\begin{array}{ll}
 \delta {\bf f}_{\sigma} &= \int_{\Delta s} \frac{\partial}{\partial s}(\sigma {\bf \tau}) ds
 (R_0/ \rho_1 v^2_0)/(R^2_0 \delta x \delta
 z)\\
 &= [(\sigma {\bf \tau})_2-(\sigma {\bf \tau})_1 ] /\rho_1 v^2_0 R_0 \delta x \delta z\\
 &= \{[(1/Ca -\Theta) {\bf \tau} ]_2-[(1/Ca -\Theta) {\bf \tau} ]_1 \} /Re  \delta x \delta z,
 \end{array}
 \end{equation}
where ${\bf \tau}$ is an unit tangent vector, $s$ is the arc length
along the interface, $\sigma(=\sigma_0+\sigma_\Theta \Theta)$ is the  surface tension coefficient
 and $Ca(=v_0 \mu_1/\sigma_0$) is the Capillary number.
The surface tension at the interface is distributed to the grid points by means of the weighting function (6).
More details of the numerical methods were presented in \cite{12}.

\section{Dimension analysis}

The system of thermocapillary droplet migration, which has the five basic dimensions
(mass $M$, length $L$, time $T$, temperature $K$ and quantity of heat $Q$),
is governed by the seven non-dimensional parameters
($Re$, $Ma$, $Ca$, $\rho_2$, $\mu_2$, $k_2$ and $\kappa_2$). They
are related to the eight independent characteristic quantities
($R_0$, $v_0$, $GR_0$, $\sigma_0$, $\rho_1$, $\mu_1$, $k_1$ and $\kappa_1$). Based on the above five basic dimensions,
 the eight characteristic quantities can be described as
 $R_0 \sim L$, $v_0 \sim L/T$, $GR_0 \sim K$, $\sigma_0 \sim M T^{-2}$, $\rho_1 \sim ML^{-3}$, $\mu_1 \sim ML^{-1}T^{-1}$,
 $k_1 \sim QL^{-1}K^{-1}$ and $\kappa_1 \sim L^2T^{-1}$. By using the dimension-analysis method,
 four time scales describing different physical processes are determined as the convective time scale ($T_c=R_0/v_0$),
 the momentum diffusion time scale ($T_m=R^2_0 \rho_1/\mu_1$),
 the thermal diffusion time scale ($T_t=R^2_0/\kappa_1$) and the capillary action time scale ($T_\sigma=R_0\mu_0/\sigma_0$).
 When the convective time scale $T_c$ is taken as the scaled time in Sect. 2, the other three dimensionless time scales are written as
 $Re=T_m/T_c$, $Ma=T_t/T_c$ and $Ca=T_\sigma/T_c$, respectively. They
 are exactly the three non-dimensional parameters of the system.

 In \cite{12}, the numerical studies on the thermocapillary droplet migration focused on the parameters of the system
  $0.66 \leq Re \leq 53.4$, $44.7 \leq Ma \leq 3622.8$ and $0.0044 \leq Ca \leq 0.040$, which are regulated with the radius $R_0$.
 For each $R_0$, $Ma$ is the largest one in the three parameters of the system.
 It is clear that the thermal diffusion time scale $T_t$ is the largest one of the system and controls
 the whole migration process. So, a steady or unsteady state as the terminal one
 of the thermocapillary droplet migration will be expected to reach in a time scale of this order, or $t = O(Ma)$ in the dimensionless terms.

\section{Results and analysis}

To verify the accuracy of the numerical model in Sect. 2,
we adopt very small Re and Ma numbers to determine the terminal migration velocity of
the droplet and compare with the analytical solution in the limits of zero Re and Ma numbers given in the Appendix. The
non-dimensional parameters are chosen as Re=0.01, Ma=0.01, Ca=0.01, $\rho_2=\mu_2=k_2=\kappa_2=0.5$.
The computational domain is fixed to the size $6 \times 12$. Based on $72 \times 144$, $144 \times 288$, $288 \times 576$ and $432 \times 864$
grid points, i.e.,12, 24, 48 and 72 grid points per droplet radius, migration velocities of the droplet are plotted
in Fig. 2(a). For each grid resolution, the migration velocity of the droplet increased from zero reaches a steady value at last
and approximates to the analytical result $V_\infty=0.22$.
A convergent trend is found when increasing the grid resolution in the simulations.
In Fig. 2(b), the migration velocities of the droplet at four Re(= 0.005,0.01,0.05 and 0.1), Ma=0.01, Ca=0.01, $\rho_2=\mu_2=k_2=\kappa_2=0.5$ with
the grid resolution for 48 grid points per droplet radius exhibit a convergent approximation to the analytical result with an error (about 9\%) when decreasing Re numbers.
In the following calculations, we fix 48 grid points per droplet radius as the
grid resolution. It is expected that thermal boundary layers around the droplet surface at large Ma numbers
have a thickness of $O(Ma^{-1/2})$\cite{15}.
So the above grid resolution can be used to depict
the thermocapillary droplet migration in a large range of Ma numbers.

The silicone oil of nominal viscosity 5cst and the FC-75 Fluorinert liquid,
i.e., the working media in the space experiments\cite{11}, are adopted as the continuous phase
fluid and the droplet, respectively.
The physical parameters
of the continuous fluid and the droplet at temperature $25^o$C are
given in Table I.
$\sigma_T$ is fixed as -0.044 dyn/cmK\cite{11}
and $\sigma_0 \approx$ 6 dyn/cm\cite{35} is adopted.
From the values of the continuous fluid parameters, Prantdl number ($Pr=Ma/Re=\mu_1/\rho_1\kappa_1$) is determined as 67.8.
 The computational domain is chosen as $\{x,z\} \in \{[0,4], [-4,44]\}$. The initial droplet is
placed at the position (0,0) and the integration time step is varied in $1 \times 10^{-5} - 2 \times 10^{-4}$
depending on the migration velocity.
To simulate the migration process in the longer time scale $O(Ma)$ at moderate and large Ma numbers,
we regulate  $R_0$ to approach the given Ma number.

\subsection{Flow field with the temperature gradient G=12K/cm}

To simulate the migration process with $G$= 12 K/cm,
the droplets with $R_0=0.05$cm, 0.075cm and 0.10cm are taken to make the systematic parameters,
 which are presented in Table II and hereinafter referred to as Ma numbers.
For $Ma \leq 178.9$, the thickness of thermal boundary layers is not smaller than 1/13.4. So
the above grid resolution is sufficiently high to
describe the thermal boundary layers for $Ma \leq 178.9$.

Figure 3(a) displays the time evolution of droplet migration velocities for Ma=44.7, 100.6 and 178.9.
The maximal computational time is $T_{max}=220$.
As given in Sect. 3, the controlling time scale of whole migration process is the relative thermal diffusion time scale
 $T_t \sim O(Ma)$. Since $T_{max}$ is beyond $T_t$,
all physical processes including the thermal diffusion
in the system are fully developed at $T_{max}$.
The whole migration processes exhibit three stages based on the curve features .
At the initial one($t \le 3$), the migration velocities markedly increase from zero to
local maximums.
Then they have decreasing-increasing processes in the range $3<t \le 70$. After the
oscillating processes, the migration velocities trend to two different qualitative behaviors in the range $70<t \le T_{max}$
depending on the Ma numbers.
One is the approaching a steady value (for Ma=44.7), the other two are the monotonically
increasing with time and do not reach any constants (for Ma=100.6 and 178.9). The increasing slope for Ma=178.9
is larger than that for Ma=100.6.
To further depict the above migration processes,
time evolution of temperature difference $\Delta \Theta[=\Theta(1,0)-\Theta(1,\pi)$] between
 the front ($\theta=0$) and  the rear ($\theta=\pi$) of the droplet is shown in Fig. 3(b).
$\Delta \Theta$ initially drops from 2 and then has an oscillating process.
At last, it approaches a constant for Ma=44.7 and monotonically increases
with time for Ma=100.6 and 178.9.
The stationary temperature difference
satisfies the requirement of steady thermocapillary migration of the droplet.
However, the increasing temperature difference
deviates from the requirement of steady thermocapillary migration of the droplet.
Thus, it is concluded that the thermocapillary droplet migration at the moderate/large Ma numbers
can/cannot reach a steady state and thus in a steady/unsteady process.

In Fig. 4(a), terminal streamlines in a reference frame
moving with the droplets  at $T_{max}$  for Ma=44.7, 100.6 and 178.9 are shown.
The droplets are located at the positions $z_c=$24.7, 24.6 and 29.2, respectively.
Two vortices symmetric about the $z$-axis appear within the droplet.
For each Ma number, the external streamlines go around the
droplet and the flow over the droplet does not separate.
Although the terminal streamlines evolved in the range of Ma numbers correspond to both the steady and unsteady states,
their patterns are similar and almost independent of Ma numbers. These properties reveal that
the instantaneous velocity fields in the unsteady migration processes at large Ma numbers may have
the forms of the steady ones at moderate Ma numbers.

Figure 4(b) displays terminal isotherms in the laboratory coordinate frame at $T_{max}$
for Ma=44.7, 100.6 and 178.9. For each Ma number,
the whole temperature field is divided into external and internal domains of the droplet.
On the one hand, the isotherms in the external domain exhibit three kinds of characteristics in the different areas.
The first one is the unperturbed uniformly-spaced parallel lines above the droplet.
The second one is the bending isotherms near the droplet to the contrary of the migration direction.
More and more isotherms are concentrated to the interface of the droplet as Ma number increases, so that the
thermal boundary layer is formed near the interface.
The last one is the curved temperature field with a gradient below the droplet.
 In the thermal boundary layer, there appear two regions related to the
thermal transfer across the interface. One is the most of interface with $\frac{\partial \Theta_i}{\partial r}(1,\theta)>0$,
 the other is a local part near the rear of the droplet with $\frac{\partial \Theta_i}{\partial r}(1,\theta)<0$.
The thermal flux across the interface in the former/latter one goes from the outside/inside of the droplet to the inside/outside.
The former/latter one is larger/smaller as Ma number increases. So, for large Ma numbers, the thermal flux
across the interface at the terminal states almost goes from the outside of the droplet to the inside.
On the other hand, the isotherms in the internal domain of the droplet exhibit different kinds of the patterns depending on Ma numbers.
For Ma=44.7, the pattern of the isotherms has a small cap-type when the droplet migrates in the steady process.
In this case, the temperature in the small cap-type isotherm ($\Theta_2=15.8$) is the lowest and about 2.8 lower than
the front temperature.
For Ma=100.6, the pea-type isotherms as the terminal state are formed in the unsteady migration process.
At this time, the temperature in the pea-type isotherm ($\Theta_2=12.8$) is the lowest and about 4.4 lower than
the front temperature.
For Ma=178.9, the isotherms with two vortices are generated when the droplet is in the accelerating process.
The lowest temperature within the droplet ($\Theta_2=11$) appears in the center of the vortex and is about 9.7 lower than the front temperature.
In general, the thermal convection and conduction are two ways of heat transfer in the system,
but the droplet can only obtain the thermal energy though the thermal conduction across the interface.
For moderate Ma numbers, the thermal convection around the droplet
and the thermal conduction across the droplet are equally important to heat transfer in the system, so that
 the external and internal temperature of the droplet
may have a fixed difference. This leads to a steady migration process.
For large Ma numbers, the thermal convection around the droplet is a stronger way of heat transfer in the system
than the thermal conduction across the droplet. The internal temperature of the droplet is more difficult to increase than
the external one of the droplet during the migration of droplet.
In the whole migration process, although the internal temperature of the droplet
increases, its increment is far lower than that of the external temperature.
The increasing difference of temperature leads to an unsteady migration precess.

Figure 4(c) displays the relative temperature distributions $\Delta \Theta/Re=[\Theta(1,\theta)- \Theta(1,\pi)]/Re$
along the interface from the front ($\theta=0$)
to the rear ($\theta=\pi$) of the droplet for the terminal states shown in Fig. 4(b).
In the range of Ma numbers, $\Delta \Theta/Re$ monotonously decreases as $\theta$ increases, i.e.,
$\frac{\partial \Delta \Theta}{\partial \theta}/Re =\frac{\partial \Theta}{\partial \theta}/Re<0$.
Although $-\frac{\partial \Theta}{\partial \theta}/Re>0$ holds on the whole surface,
its averaged values $|\frac{\Delta \Theta}{\Delta \theta}|_t/Re \approx [\Theta(1,0)- \Theta(1,\pi/2)]/(\pi Re/2)$
on the top surface \{$\theta \in [0,\frac{\pi}{2}]$\}
and $|\frac{\Delta \Theta}{\Delta \theta}|_b/Re \approx [\Theta(1,\pi/2)- \Theta(1,\pi)]/(\pi Re/2)$ on the bottom  surface \{$\theta \in (\frac{\pi}{2},\pi]$\}
are varied depending on Ma numbers.
For Ma=44.7, $|\frac{\Delta \Theta}{\Delta \theta}|_t/Re < |\frac{\Delta \Theta}{\Delta \theta}|_b/Re$.
For Ma=100.6, the above relation holds, but their difference decreases.
For Ma=178.9, $|\frac{\Delta \Theta}{\Delta \theta}|_t/Re \approx |\frac{\Delta \Theta}{\Delta \theta}|_b/Re$,
and an inflection point ($\frac{\partial^2 \Theta}{\partial \theta^2}=0$) appears near $\theta=\pi/2$.
The net force acting on the droplet exerted by the continuous phase fluid in the vertical direction\cite{1} is written as
\begin{equation}
\begin{array}{ll}
F_z & = \int_S  {\bf n} \cdot {\bf \Pi}_1 \cdot {\bf i}_z dS \\
    & = \int_0^\pi ( \Pi_{1nn} \cos \theta - \Pi_{1n\tau} \sin \theta ) d\theta\\
    & = \int_0^\pi [ \Pi_{1nn} \cos \theta - (\Pi_{2n\tau} - \frac{1}{Re Ca} \frac{\partial \sigma}{\partial \theta} )\sin \theta] d\theta\\
    & = \int_0^\pi ( \Pi_{1nn} \cos \theta - \Pi_{2n\tau} \sin \theta - \frac{1}{Re} \frac{\partial \Theta}{\partial \theta} \sin \theta) d\theta\\
    & \approx -\frac{2\pi }{Re}(1+\mu_2) V_\infty - \frac{1}{Re} \int_0^\pi \frac{\partial \Theta}{\partial \theta} \sin \theta d\theta\\
& \approx -\frac{2\pi }{Re}(1+\mu_2) V_\infty + \frac{1}{Re} \Sigma_{i=1}^{\frac{N}{2}} |\frac{\partial \Theta}{\partial \theta}|_t \sin \theta_i \Delta \theta
+ \frac{1}{Re} \Sigma_{i=\frac{N}{2}+1}^{N} |\frac{\partial \Theta}{\partial \theta}|_b \sin \theta_i \Delta \theta \\
& = -\frac{2\pi }{Re}(1+\mu_2) V_\infty +\frac{1}{Re} (|\frac{\Delta \Theta}{\Delta \theta}|_t + |\frac{\Delta \Theta}{\Delta \theta}|_b) \Delta \theta \Sigma_{i=1}^{\frac{N}{2}} \sin \theta_i,
\end{array}
\end{equation}
where ${\bf n}$ is the unit normal to the surface,
${\bf \Pi}_1= -p_1 {\bf I} + \frac{1}{Re} [\nabla {\bf v}_1 +(\nabla {\bf v}_1)^T]$ is
the stress tensor of the continuous phase and $V_\infty$ is the instantaneous migration velocity of the droplet.
In the above derivation, the tangential stress balance
$\Pi_{1n\tau}-\Pi_{2n\tau}= - \frac{1}{Re Ca} \frac{\partial \sigma}{\partial \theta}$
at the interface of the two-phase fluids is used. Due to the above investigations in the steady and unsteady migration processes,
the instantaneous velocity fields of two-phase fluids are assumed as those in the limits of zero Re and zero Ma numbers given in the Appendix.
This assumption is also supported with
the same velocity fields at the zero-order approximation of thermocapillary migration of a spherical droplet at zero Ma(Re),
small Ma(Re) and large Ma(Re) numbers\cite{2,3,8}.

In Eq. (9), the net force includes of the first term for the drag force and the second term for the driving force.
When the driving force is larger than the drag force, i.e., $F_z>0$, the droplet migrates in an accelerating process.
When the driving force is equal to the drag force, i.e., $F_z=0$, the droplet is in a steady state.
The surface tensions of the top and bottom surfaces of the droplet simultaneously contribute to the driving force. However,
for moderate Ma numbers, the surface tension
of the bottom surface, which is larger than that of the top surface, is the main component
of the driving force. For large Ma numbers,
the surface tensions of the top and bottom surfaces of the droplet have the same order of magnitude
 in the driving force when the inflection point appears.
The driving force on the droplet decreases as Ma number increases.

\subsection{Flow field with the temperature gradient G = 9 K/cm}

To simulate the migration process with $G$=9K/cm,
the droplets with $R_0=0.075$cm, 0.10cm and 0.125cm are taken to make the systematic parameters,
 which are presented in Table III.
For $Ma \leq 209.7$, the thickness of thermal boundary layer is not smaller than 1/14.5. So
the above grid resolution is sufficiently high to
describe the thermal boundary layers for $Ma \leq 209.7$.

Figure 5(a) displays the time evolution of droplet migration velocities for Ma=75.5, 134.2 and 209.7.
The maximal computational time $T_{max}=220$
is beyond the relative thermal diffusion time scale $T_t \sim O(Ma)$.
The initial migration processes exhibit increasing-decreasing-increasing processes in the range $0<t \le 100$. Then
the migration velocities trend to two different qualitative behaviors in the range $100<t \le T_{max}$
depending on Ma numbers.
One is the approaching a steady value (for Ma=75.5), the other two are the monotonically
increasing with time and do not reach any constants (for Ma=134.2 and 209.7). The increasing slope for Ma=209.7
is larger than that for Ma=134.2.
These migration processes can be understood in the time evolution of temperature difference $\Delta \Theta$ between
 the front and the rear of the droplet shown in Fig. 5(b).
For Ma=75.5,  the stationary temperature difference
satisfies the requirement of steady thermocapillary migration of the droplet.
However, for Ma=134.2 and 209.7, the monotonically increasing temperature difference
deviates from the requirement of steady thermocapillary migration of the droplet.
Thus, it can be concluded that the thermocapillary droplet migration at moderate/large Ma numbers
can/cannot reach a steady state and thus in a steady/unsteady process.

In Fig. 6(a), terminal streamlines in a reference frame
moving with the droplets at $T_{max}$ for Ma=75.5, 134.2 and 209.7 are shown.
The droplets are located at the positions $z_c=$24.2, 26.3 and 34.3, respectively.
The terminal streamlines evolved in the range of Ma numbers reveal that
the instantaneous velocity fields are similar no matter they are the steady states or the unsteady ones.
Figure 6(b) displays terminal isotherms in the laboratory coordinate frame at $T_{max}$
for Ma=75.5, 134.2 and 209.7. For each Ma number,
the isotherms in the external domain of the droplet bend to the contrary of the migration direction and converge near
the interface of the droplet to form a thermal boundary layer.
 In the thermal boundary layer, there appear two regions with $\frac{\partial \Theta_i}{\partial r}(1,\theta)>0$
 for the most of interface and $\frac{\partial \Theta_i}{\partial r}(1,\theta)<0$ for
 a local part near the rear of the droplet. They exhibit two thermal fluxes with different directions across the interface
 and evolve as Ma number increases. For large Ma numbers, the thermal flux
across the interface at the terminal states almost goes from the outside of the droplet to the inside.
Meanwhile, the isotherms in the internal domain of the droplet exhibit different kinds of patterns depending on Ma numbers.
For Ma=75.5, the pattern of the isotherms has a small cap-type when the droplet migrates in the steady process.
The lowest temperature within the droplet is $\Theta_2=13.8$ in the small cap-type isotherm and about 3.5 lower than
the front temperature.
The constant temperature difference $\Delta \Theta=3.5$ between the outside and inside of the droplet
implies a steady migration precess. It is based on the fact that the thermal convection and conduction
are two equally important to heat transfer in the system at moderate Ma numbers.
For Ma=134.2, the pea-type isotherms are formed in the unsteady migration process.
The lowest temperature within the droplet is $\Theta_2=12$ in the pea-type isotherm and about 6.2 lower than the front temperature.
For Ma=209.7, the isotherms with two vortices are generated when the droplet is in the accelerating process.
The lowest temperature within the droplet is $\Theta_2=11.2$ in the center of the vortex and about 15.0 lower than the front temperature.
The temperature difference $\Delta \Theta$ between the outside and inside of the droplet increases
from 6.2 to 15 when Ma number increases from 134.2 to 209.7. This leads to an unsteady migration precess.
The reason is that the thermal convection around the droplet is a stronger way of the heat transfer in the system
than the thermal conduction across the droplet at large Ma numbers.
As shown in Fig. 6(c),
$\Delta \Theta/Re$ in the whole surface monotonously decreases as $\theta$ increases in the range of Ma numbers.
For Ma=75.5, $|\frac{\Delta \Theta}{\Delta \theta}|_t/Re < |\frac{\Delta \Theta}{\Delta \theta}|_b/Re$.
For Ma=134.2, the above relation holds, but their difference decreases.
The surface tension of the bottom surface, which is larger than that of the top surface, is the main component
of the driving force in Eq. (9).
For Ma=209.7, $|\frac{\Delta \Theta}{\Delta \theta}|_t/Re > |\frac{\Delta \Theta}{\Delta \theta}|_b/Re$,
and an inflection point ($\frac{\partial^2 \Theta}{\partial \theta^2}=0$) appears near $\theta=\pi/2$.
The surface tension of the top surface, which is larger than that of the bottom surface,
is the main component of the driving force in Eq. (9) after the inflection points appears.
The driving force on the droplet
decreases first and then increases as Ma number increases.
From Figs. 4(c) and 6(c), it can be concluded that the appearance of the inflection point implies that
the driving force will increase as Ma number increases.

\section{Conclusions and discussions}
\label{sec:sum}
To sum up, thermocapillary migration of a planar droplet at moderate and large Ma numbers has been
investigated analytically and numerically.
Firstly, by using the dimension-analysis method, the thermal diffusion time scale
has been determined as the controlling one of the thermocapillary droplet migration system.
During this time, the whole thermocapillary migration process is fully developed.
Then, by the aid of the front-tracking method, the numerical simulations in a longer time scale,
which is beyond the thermal diffusion time scale, have exhibited that
the terminal states of thermocapillary droplet migration at moderate and large Ma numbers are steady and unsteady, respectively.
In the terminal states, the instantaneous velocity fields in the unsteady migration processes at large Ma numbers
have the forms of the steady ones at moderate Ma numbers.
However, in view of the former instantaneous temperature fields, the surface tension of the top surface of the droplet
 gradually becomes the main component of the driving force on the droplet after the inflection point appears.
 It is different from that the surface tension
 of the bottom surface of the droplet is the main component of the driving force on the droplet for the latter ones.
The analysis based
on the time evolution of velocity and temperature fields implies
that the temperature difference between the front and rear of the droplet
is a constant for moderate Ma numbers and has
the increasing tread for large Ma numbers.
It satisfies/deviates from the requirement of steady thermocapillary migration of the droplet at moderate/large Ma numbers.
These phenomena originate from the evolution of the heat transfer across/around the droplet in the system
depending on Ma numbers. The significance of the thermal
 convection around the droplet is higher than/just as the thermal conduction across the droplet at large/moderate Ma numbers.

\newpage
\textbf{Acknowledgments}
This work was partially supported by the National Science
Foundation through the Grants No. 11172310 and No. 11472284.
The author thanks the  research computing facility
of the Institute of Mechanics of the Chinese Academy of Sciences for
assisting in the computation.

\newpage
\textbf{Appendix: A general solution of steady thermocapillary
migration of a planar droplet in the limits of zero Re and zero Ma numbers}

The steady incompressible continuous, momentum and energy equations for the continuous phase fluid and the
droplet in the limits of zero Re and Ma numbers can be
written in the following dimensionless form in a polar
coordinate system ($r,\theta$) moving with the droplet velocity $V_{\infty}$
\begin{equation}
\begin{array}{l}
\nabla \cdot {\bf v}_i=0,\\
\nabla p_i= \frac{\mu_i}{Re} \Delta {\bf v}_i,\\
\Delta \Theta_i=0,
\end{array}
\end{equation}
where ${\bf v}_i=(v_{ir},v_{i\theta})$.
The boundary conditions are written in the form of dimensionless as
\begin{equation}
(v_{1r}, v_{1\theta}) \to (-V_\infty \cos \theta, V_\infty \sin \theta), \ \ \
p_1 \to 0, \ \ \ \Theta_1 \to r \cos \theta
\end{equation}
 at infinity and
\begin{equation}
\begin{array}{l}
v_{1r}(1,\theta) =v_{2r}(1,\theta)=0,\\
v_{1\theta}(1,\theta) =v_{2\theta}(1,\theta),\\
{\bf n} \cdot {\bf \Pi}_1 - {\bf n} \cdot {\bf \Pi}_2=\frac{1}{ReCa} [\sigma {\bf n}- \nabla \sigma],\\
\Theta_1(1,\theta) =\Theta_2(1,\theta),\\
\frac{\partial \Theta_1}{\partial r}(1,\theta) =k_2 \frac{\partial \Theta_2}{\partial r}(1,\theta),
\end{array}
\end{equation}
at the interface of the two-phase fluids,
${\bf n}$ is the unit vector normal to the interface,
${\bf \Pi}_i= -p_i {\bf I} + \frac{\mu_i}{Re} [\nabla {\bf v}_i +(\nabla {\bf v}_i)^T]$ is
the stress tensors of the two-phase fluids.
Following the methods for solving the linear model\cite{2,31}, the
solutions of the equations (10) with the boundary conditions (11)(12) can be determined as
\begin{equation}
\begin{array}{l}
\Psi_1(r,\theta)=-V_\infty(r-\frac{1}{r}) \sin \theta,\\
\Psi_2(r,\theta)=-V_\infty r(r^2-1) \sin\theta,\\
p_1(r,\theta)=0,\\
p_2(r,\theta)=p'_0 -8\mu_2 V_\infty r \cos \theta/Re,\\
\Theta_1(r,\theta)=(r+\frac{1-k_2}{1+k_2}\frac{1}{r}) \cos \theta,\\
\Theta_2(r,\theta)=\frac{2}{1+k_2}r \cos \theta,
\end{array}
\end{equation}
where the velocity fields $(v_{ir}, v_{i\theta})=(\frac{\partial \Psi_i}{r\partial \theta}, -\frac{\partial \Psi_i}{\partial r})$
 are written in terms of the stream functions $\Psi_i(r,\theta)$ of the two-phase fluids.
The pressure fields $p_i(r,\theta)$ with a constant $p'_0$ are obtained
by integrating the momentum equations in Eqs. (10).

In addition to the above boundary conditions (12) at the interface,
 the steady thermocapillary droplet migration requires that the
total net force acting on the droplet is zero. In particular, the zero net force  in the vertical direction\cite{1} is expressed as
\begin{equation}
 \int_{S_1} (\Pi_{1n\tau} \sin \theta -\Pi_{1nn} \cos \theta) dS =\int_0^{\pi} (\Pi_{1n\tau} \sin \theta -\Pi_{1nn} \cos \theta)|_{r=1} d\theta =0.
\end{equation}
When the shear stress $\Pi_{1r\theta}$ for the continuous fluid is replaced by using the stress boundary condition in Eqs. (12), the steady droplet migration speed
is derived as
\begin{equation}
V_\infty=-\frac{1}{2 \pi (1+\mu_2)} \int_0^\pi \frac{\partial \Theta_1}{\partial \theta}(1,\theta) \sin \theta d \theta=
\frac{1}{2(1+\mu_2)(1+k_2)}.
\end{equation}

\newpage
Table I. Physical parameters of the continuous phase fluid (5cst
Silicone oil) and the droplet (FC-75) at temperature $25^o$C,
which are the working media in the space experiment\cite{11}.

\begin{tabular}{l|llll}
 \hline
        &    $\rho$($g/cm^3$) & $\mu$($10^{-2}dyn s/cm^2$) & $k$($W/mK$) & $\kappa$($10^{-4}cm^2/s$)
\\ \hline
Silicone oil &  0.91      & 4.268 &0.111 & 6.915 \\
FC-75        &  1.77      & 1.416 &0.063 & 2.018 \\
\hline
\end{tabular}

Table II. Correspondence of non-dimensional parameters Re, Ma and
Ca to the droplet radius $R_0$ for the thermocapillary droplet migration in a flow
field with the temperature gradient $G=12K/cm$.

\begin{tabular}{l|lrl}
 \hline
     $R_0(cm)$ &      Re    &   Ma   &  Ca
\\ \hline
0.05         &  0.66        & 44.7    & 0.0044\\
0.075        &  1.48        & 100.6   & 0.0066\\
0.10         &  2.64        &  178.9  & 0.0088\\
\hline
\end{tabular}

Table III. Correspondence of non-dimensional parameters Re, Ma and
Ca to the droplet radius $R_0$ for the thermocapillary droplet migration in a flow
field with the temperature gradient $G=9K/cm$.

\begin{tabular}{l|lrl}
 \hline
     $R_0(cm)$ &      Re    &   Ma   &  Ca
\\ \hline
0.075        &  1.11        & 75.5   & 0.0050\\
0.10         &  1.98        &  134.2  & 0.0066\\
0.125        &  3.09        &  209.7  & 0.0082\\
\hline
\end{tabular}

\newpage
\textbf{Figure caption}

Fig.~1. Schematic of the computational domain for thermocapillary migration of a planar droplet.
The top, bottom and right walls are non-slip boundaries. The $z$-axis is the mirror
symmetric axis of the system.

Fig.~2. Comparison between the numerical results at Ma=0.01, Ca=0.01
and $\rho_2=\mu_2=k_2=\kappa_2=0.5$
and the analytical one of droplet migration velocity in the limits of zero Re
and Ma numbers. (a) the numerical results at a fixed Re=0.01
with four grid resolutions; (b) the numerical results at four Re(=0.005, 0.01, 0.05 and 0.1)
with a fixed grid resolution for 48 grid points per droplet radius.

Fig.~3. (a) Droplet migration velocities in a flow field with the temperature
gradient $G=12 K/cm$ versus non-dimensional time for Ma=44.7, 100.6 and 178.9;
(b) Time evolution of temperature difference between the front
and the rear of the droplet.

Fig.~4. (a) Terminal streamlines in a reference frame moving with the droplet at $T_{max}=220$
for Ma=44.7, 100.6 and 178.9; (b) Terminal isotherms in a laboratory coordinate frame;
(c) Relative temperature distributions  along the interface from the front to the rear of the droplet
for the terminal states in (b).

Fig.~5. (a) Droplet migration velocities in a flow field with the temperature
gradient $G=9 K/cm$ versus non-dimensional time for Ma=75.5, 134.2 and 209.7;
(b) Time evolution of temperature difference between the front
and the rear of the droplet.

Fig.~6. (a) Terminal streamlines in a reference frame moving with the droplet at $T_{max}=220$
for Ma=75.5, 134.2 and 209.7; (b) Terminal isotherms in a laboratory coordinate frame;
(c) Relative temperature distributions along the interface from the front  to the rear of the droplet
for the terminal states in (b).


\newpage
\begin{thebibliography}{}
\bibitem{1}
R. S. Subramanian and R. Balasubramaniam, The motion of bubbles and drops in reduced gravity,
Cambridge, England: Cambridge University Press, 2001.

\bibitem{2}
N. O. Young, J. S. Goldstein and M. J. Block, The motion of bubbles
in a vertical temperature gradient, {\it J. Fluid Mech.} {\bf 6},
(1959) p. 350.

\bibitem{3}
R. S. Subramanian, Thermocapillary migration of bubbles and droplets,
{\it Adv. Spcae Res.} {\bf 3}, (1983) p. 145.

\bibitem{4}
B. Braun, C. Ikier and H. Klein,  Thermocapillary migration of
droplets in a binary mixture with miscibility gap during
liquid/liquid phase separation under reduced gravity, {\it J.
Colloid Interface Sci.} {\bf 159}, (1993) p. 515.

\bibitem{5}
R. Balasubramaniam and A.-T. Chai, Thermocapillary migration of
droplets: an exact solution for small Marangoni numbers, {\it J.
Colloid Interface Sci.} {\bf 119}, (1987) p. 531.

\bibitem{6}
H. Haj-Harir, Q. Shi and A. Borhan, Thermocapillary motion of
deformable drops at finite Renoylds and Marangoni numbers, {\it
Phys. Fluids}, {\bf 9} (1997), p. 845.

\bibitem{7}
R. Balasubramaniam, C. E. Lacy, G. Woniak and R. S. Subramanian,
Thermocapillary migration of bubbles and drops at moderate values
of Marangoni numbers in reduced gravity, {\it Phys. Fluids} {\bf
8} (1996) p. 872.

\bibitem{8}
R. Balasubramaniam and R. S. Subramanian, The migration of a drop in
a uniform temperature gradient at large Marangoni numbers, {\it
Phys. Fluids} {\bf 12}, (2000) p. 733.

\bibitem{9}
X. Ma, R. Balasubramiam and R. S. Subramanian,  Numerical simulation
of thermocapillary drop motion with internal circulation, {\it
Numer. Heat Transfer}, A{\bf 35}, (1999) p. 291.

\bibitem{10}
P. H. Hadland, R. Balasubramaniam and G. Wozniak, Thermocapillary
migration of bubbles and drops at moderate to large Marangoni
number and moderate Reynolds number in reduced gravity, {\it
Exp.  Fluid}, {\bf 26}, (1999) p. 240.

\bibitem{11}
J. C. Xie, H. Lin, P. Zhang, F. Liu and W. R. Hu,  Experimental
investigation on thermocapillary drop migration at large Marangoni
number in reduced gravity, {\it J. Colloid Interface Sci.} {\bf
285}, (2005) p. 737.

\bibitem{a1}
V. Ludviksson and E. N. Lightfoot, The dynamics of thin liquid films
in the presence of surface-tension griendients, {\it AiChE J.} {\bf 17},
(1971) p. 1166.

\bibitem{a7}
G. F. Teletzke, H. T. Davis and L. E. Scriven, How liquids spread on
solids, {\it Chem. Engin. Commun.} {\bf 55} (1987) p. 41.

\bibitem{a2}
Y. J. Zhao, F. J. Liu and C. H. Chen, Thermocapillary actuation
of binary drops on solid surfaces, {\it Appl. Phys. Lett.}
{\bf 99} (2011) p. 104101.

\bibitem{a3}
D. E. Kataoka and S. M. Troian, Stabilizing the advancing front
of thermally driven climbing films, {\it J. Colloid Interface Sci.}
{\bf 203} (1998) p. 335.

\bibitem{a4}
V. Pratap, N. Moumen and R. S. Subramanian, Thermocapillary motion
of a liquid drop on a horizontal solid surface, {\it Langmuir}
{\bf 24} (2008) p. 5185.

\bibitem{a5}
J. M. Gomba and G. M. Homsy, Regimes of thermocapillary migration
of droplets under partial wetting conditions, {\it J. Fluid Mech.}
{\bf 647} (2010) p. 125.

\bibitem{a6}
M. L. Ford and A. Nadim, Thermocapillary migration of an attached
drop on a solid surface, {\it Phys. Fluids} {\bf 6} (1994) p. 3183.


\bibitem{12}
Z.-B. Wu and W. R. Hu, Thermocapillary migration of a planar droplet
at moderate and large Marangoni numbers. {\it Acta Mech.}, {\bf 223}
(2012) p. 609.

\bibitem{13}
G. Tryggvason, et al, A front-tracking method for the computations
of multiphase flow, {\it J. Comput. Phys.} {\bf 169}, (2001) p. 708.


\bibitem{14}
C. S. Peskin, Numerical analysis of blood flow in the heart, {\it
J. Comput. Phys.} {\bf 25}, (1977) p. 220.

\bibitem{15}
R. Balssubramaniam and R. S. Subramanian, Thermocapillary bubble
migration-thermal boundary layers for large Marangoni numbers, {\it
Int. J. Multiphase Flow} {\bf 22}, (1996) p. 593.

\bibitem{35}
S. Someya and T. Munakata, Measurement of the interface tension of
immiscible liquids interface. {\it J. Crystal Growth} {\bf 275},
(2005) c343.

\bibitem{31}
J. Happel and H. Brenner, {\it Low Reynolds number hydrodynamics},
 Prentice-Hall, Englewood Cliffs, N. J. (1965).



\end{thebibliography}
\end{document}